# VELOCITY CORRELATIONS OF GALAXY CLUSTERS

Renyue Cen[1], Neta A. Bahcall[1] and Mirt Gramann[1,2]

email: cen@astro.princeton.edu

[1] Princeton University Observatory, Princeton, NJ 08544
[2] Tartu Astrophysical Observatory, Tõravere, Estonia

# ABSTRACT

We determine the velocity correlation function, pairwise peculiar velocity difference, and root-mean-square pairwise peculiar velocity dispersion of rich clusters of galaxies, as a function of pair separation, for three cosmological models: $\Omega = 1$ and $\Omega = 0.3$ CDM, and $\Omega = 0.3$ PBI models (all flat and COBE-normalized). We find that close cluster pairs, with separation $r \leq 10h^{-1}$Mpc, exhibit strong *attractive* peculiar velocities in all models; the cluster pairwise velocities depend sensitively on the model. The mean pairwise attractive velocity of clusters on $5h^{-1}$Mpc scale ranges from $\sim 1700$km s$^{-1}$ for $\Omega = 1$ CDM, to $\sim 1000$km s$^{-1}$ for PBI, to $\sim 700$km s$^{-1}$ for $\Omega = 0.3$ CDM. The small-scale pairwise velocities depend also on cluster mass: richer, more massive clusters exhibit stronger attractive velocities than less massive clusters. On large scales, from $\sim 20$ to $200h^{-1}$Mpc, the cluster peculiar velocities are increasingly dominated by bulk and random motions; they are independent of cluster mass. The cluster velocity correlation function, which reflects the bulk motion minus the relative motion of pairs, is negative on small scales for $\Omega = 1$ and $\Omega = 0.3$ CDM, and positive for PBI; this indicates stronger pairwise motion than bulk motion on small scales for CDM, and relatively larger bulk motions for PBI. The cluster velocity correlation function is positive on very large scales, from $r \sim 10h^{-1}$Mpc to $\sim 200h^{-1}$Mpc, for all models. These positive correlations, which decrease monotonically with scale, indicate significant bulk motions of clusters up to $\sim 200h^{-1}$Mpc. The strong dependence of the cluster velocity functions on models, especially at small separations, makes them useful tools in constraining cosmological models when compared with observations.

Cosmology: large-scale structure of Universe – Cosmology: theory – galaxies: clustering – galaxies: clusters of

## 1. INTRODUCTION

Observations of large-scale motions in the universe indicate that the Hubble expansion is considerably perturbed on scales up to at least $60h^{-1}$Mpc (Rubin et al. 1976; Collins et al. 1986; Dressler et al. 1987; Burstein et al. 1987; Sutherland 1988; Lynden-Bell *et al.*1988; Faber et al. 1989; Mould et al. 1991; Strauss *et al.*1992; Mathewson et al. 1992). The perturbations are believed to be caused by the gravitational growth of cosmic structures. The observed peculiar motions can therefore place critical constraints on cosmological models. One of the most important quantitative statistical tools used in contrasting models with observations is the peculiar velocity correlation function: comparing the strength and the scale on which galaxy pair motions are correlated with each other (Peebles 1987, 1993; Górski *et al.*1989; Groth et al. 1989). Almost all models and observations so far have concentrated on studying the peculiar velocities of galaxies. However, uncertainties in the observational determination of galaxy peculiar velocities, which are proportional to the galaxy distances, do not allow accurate velocity determinations except for the nearest galaxies.

The velocities of rich clusters of galaxies can be determined more accurately than the velocities of individual galaxies since the distance to each cluster can be obtained by averaging the measured distances of a large number of member galaxies. The Brightest Cluster Galaxy (BCG) distance method also appears to be promising (Lauer & Postman 1994). In addition, cluster peculiar velocities can, in principle, be determined to large distances using another, independent method of the Sunyaev-Zel'dovich (1980) effect: measuring the Doppler shift of the scattered cosmic microwave radiation (CMR) caused by the peculiar motion of the cluster relative to the CMR frame. While the effect is small, current observations are beginning to use this method to measure cluster motions. Cluster motions could therefore provide a major (and economical) tool in probing the large-scale peculiar velocity field (Bahcall, Gramann & Cen 1994; Strauss *et al.* 1994).

In this letter we determine three quantitative statistics of the peculiar motions of rich clusters of galaxies in different cosmologies: 1) the velocity correlation function of clusters; 2) the pairwise velocity difference of clusters as a function of cluster pair separation; and 3) the root-mean-square pairwise velocity dispersion of clusters as a function of pair separation. We also investigate the dependence of the above functions on cluster richness.

## 2. MODEL SIMULATIONS

A Particle-Mesh code (Cen 1992) with a box size of $800h^{-1}$Mpc is used to simulate the evolution of dark matter in different cosmological models. A large simulation box is needed in order to 1) ensure that contributions to velocities from waves larger than the box size are small, and 2) minimize uncertainties due to fluctuations in the small number of large waves. Each simulation has $500^3$ cells and $250^3 = 10^{7.2}$ particles. The nominal spatial resolution is $1.6h^{-1}$Mpc.

Three cosmological model simulations are performed: Standard Cold Dark-Matter (SCDM) model ($\Omega = 1$, $h = 0.5$, $\sigma_8 = 1.05$), low-density CDM (LCDM) model ($\Omega = 0.3$, $\lambda = 0.7$, $h = 0.67$, $\sigma_8 = 0.67$), and Primeval Baryonic Isocurvature (PBI) model ($\Omega = 0.3$, $\lambda = 0.7$, $h = 0.5$, $\sigma_8 = 1.02$). The models are normalized to the CMR anisotropy measured by COBE (Smoot et al. 1992).

Clusters are selected in each simulation using an adaptive, friends-of-friends linkage algorithm described by Suto, Cen & Ostriker (1992) and Bahcall & Cen (1992). The cluster mass threshold, within $r = 1.5h^{-1}$Mpc of the cluster center, is selected to correspond to a number density of clusters comparable to the observed density of rich ($R \geq 1$) clusters, $n_{cl} \sim 6 \times 10^{-6} h^3$Mpc$^{-3}$ (Bahcall & Cen 1993). A total of $\sim 3000$ rich clusters of galaxies are obtained in each of the $800h^{-1}$Mpc simulation models. The three-dimensional peculiar velocity of each cluster, relative to the co-moving cosmic background frame, is determined from the simulation; these velocities are used in the analyses described below. In order to investigate the dependence of the results on cluster richness (i.e., mass), we carry out the velocity analysis independently for rich ($R \geq 1$) clusters, for the less massive $R \geq 0$ clusters with a mean density of $n_{cl}(R \geq 0) \approx 1.3 \times 10^{-5} h^3$Mpc$^{-3}$, and for lower threshold groups with $n_{gr} \approx 10^{-4} h^3$Mpc$^{-3}$ (Bahcal & Cen 1993);

these comparisons are discussed in sections 3 and 4.

The sensitivity of the results to the resolution of the simulation is tested by comparing simulations with 0.8 and $1.6h^{-1}$Mpc nominal resolutions (both using the same $400h^{-1}$Mpc box). The resulting cluster velocity distributions are consistent to better than 60km s$^{-1}$. The simulation results are also compared with expectations from linear theory (Gramann et al. 1994).

## 3. VELOCITY CORRELATION FUNCTION OF CLUSTERS

The velocity correlation function of rich clusters, $\Psi_v(r) = \text{sign}| < \vec{v}_1(\vec{r}_1) \cdot \vec{v}_2(\vec{r}_1 + \vec{r}) > |^{1/2} = \text{sign}\frac{1}{2}| < [\vec{v}_1(\vec{r}_1) + \vec{v}_2(\vec{r}_1 + \vec{r})]^2 - [\vec{v}_1(\vec{r}_1) - \vec{v}_2(\vec{r}_1 + \vec{r})]^2 > |^{1/2}$ (Peebles 1987, 1993; Groth et al. 1989), is determined as a function of cluster pair separation, $r$, for all models. $\Psi_v(r)$ measures the difference of two terms — the bulk motion less the relative motion of cluster pairs. $\Psi_v$ is positive if clusters in the pairs move in the same direction, and is negative if they move in opposite directions. A random distribution of cluster motions yields $\Psi_v \sim 0$; bulk-motion exhibits $\Psi_v > 0$, and attracting or repulsing cluster pairs have $\Psi_v < 0$.

Figure 1 shows $\Psi_v(r)$ for rich ($R \geq 1$) clusters of galaxies. It reveals negative correlations for the two CDM models and positive correlations for PBI on small scales ($r < 10h^{-1}$Mpc), and positive correlations on larger scales (from $r \sim 10h^{-1}$Mpc to $\sim 200h^{-1}$Mpc) for all models. The small scale negative correlations are strongest in SCDM, with $\Psi_v(5h^{-1}Mpc) \sim -700$km s$^{-1}$. On large scales, $r \geq 20h^{-1}$Mpc, the positive correlation function decreases monotonically with scale. Table 1 summarizes the rich cluster velocity correlations.

What are the implications of the above velocity correlation function? The negative correlations on small scales ($r < 10h^{-1}$Mpc) imply that the relative motion of close rich cluster pairs dominates over their bulk motion. As we show in §4.1, close cluster pairs mostly approach each other, as expected. The cluster attractive velocities are largest in SCDM. The PBI model yields positive correlations even at these small separations, suggesting that, in addition to the cluster attractive motions (§4.1), large bulk motions (due to the large-scale bump in the power spectrum of PBI) contribute significantly (positively) to $\Psi_v > 0$. On larger scales, $r \geq 10h^{-1}$Mpc, the cluster velocity correlations become positive, and remain positive to very large scales ($r \geq 200h^{-1}$Mpc). This reflects the transition from the local gravitational attraction regime for close cluster pairs to the large scale regime where the velocity correlations are dominated by bulk motions. The PBI model exhibits the strongest positive velocity correlations on scales up to $r \sim 10^2 h^{-1}$Mpc, where it drops rapidly due to its steep slope of the power spectrum on large scales ($r \geq 400h^{-1}$Mpc). The positive correlations decrease monotonically with scale for all models for $r \geq 20h^{-1}$Mpc as expected for large-scale bulk motions. The results are consistent with the detailed large-scale cluster motions studies by Gramann et al. (1994) for these models. Comparison with expectation from linear theory on large scales shows consistency with the simulations of the low-density models, and a slight overestimate of the simulated bulk-flow of SCDM by $\sim 80$km s$^{-1}$ (Gramann et al. 1994). The positive cluster

Figure 1. Velocity correlation function of rich ($R \geq 1$) clusters of galaxies for the three models (§3).

velocity correlations on large scales suggest a similar behavior for the galaxies.

The velocity correlation function on small scales depends on cluster mass, or richness. We find that poorer, less massive clusters exhibit somewhat weaker attractive correlations on small scales than massive clusters; the large scale velocity correlations, however, are independent of cluster mass. The richness dependence of the cluster velocity correlation function is presented in Figure 2, where we compare the velocity correlation function of the rich $R \geq 1$ clusters, with that of poorer $R \geq 0$ clusters, and of groups of galaxies (see §2). The results are presented for SCDM and LCDM; PBI yields trends similar to those of LCDM. Since the small scale correlations are mostly due to attractive forces between close cluster pairs, we find that massive clusters attract each other more strongly than less massive clusters. The same trend is observed in the cluster spatial correlation function where more massive rich clusters are correlated more strongly than poorer clusters (Bahcall & Soneira 1983; Bahcall & West 1992).

Figure 2. Dependence of the cluster velocity correlation function on cluster richness ($R \geq 1$, $R \geq 0$ clusters and groups), for $\Omega = 1$ (top) and $\Omega = 0.3$ (bottom) CDM. (The large-scale $\Omega = 1$ CDM velocities are overestimated in comparison with linear theory by $\sim 80$km s$^{-1}$. (§3).

## 4. PAIRWISE VELOCITIES OF CLUSTERS
### 4.1 Velocity Difference of Cluster Pairs

The mean peculiar velocity difference of cluster pairs along the radius vector $\vec{r} = \vec{r_2} - \vec{r_1}$ joining the pair, $v_{12}(r) = -\langle(\vec{v_2} - \vec{v_1}) \cdot \vec{r}/r\rangle$, is shown in Figure 3; the brackets indicate an average over all pairs at the same separation. The pairwise peculiar velocity difference $v_{12}(r)$ is positive for attracting pairs, it is negative for repelling pairs, and it is zero for a random distribution of peculiar velocities. Bulk-flow motions do not contribute to $v_{12}$. The results show large positive pairwise velocities at small separations, $v_{12}(r \leq 10h^{-1}\text{Mpc}) \sim 500$ to $1700\text{km s}^{-1}$, which decrease monotonically to zero on large scales. All models exhibit the same behavior but with different amplitudes (Figure 3).

The strong mean pairwise cluster velocities for $r \leq 10h^{-1}\text{Mpc}$ implies a strong mean attraction of cluster pairs on these scales. The attraction is strongest for SCDM, with $v_{12}(r \sim 5h^{-1}\text{Mpc}) \sim 1700\text{km s}^{-1}$, and weakest for LCDM, with $v_{12}(r \sim 5h^{-1}\text{Mpc}) \sim 700\text{km s}^{-1}$. The PBI model yields intermediate values. The results are summarized in Table 1. The strong attractive motions in SCDM are partly due to the larger cluster masses in this model (larger than observed, and larger than LCDM and PBI; Bahcall & Cen 1992).

The monotonic decrease of $v_{12}(r)$ with $r$ reflects the relative decrease of the cluster gravitational attraction and the increased contribution of random and bulk velocities at large scales. The random and bulk velocities finally dominate on large scales, $r \geq 50-100h^{-1}\text{Mpc}$, where $v_{12} \sim 0$. The amplitude of the cluster pairwise velocity function on small scales depends on cluster richness, or mass, similar to the richness dependence of the cluster velocity correlation function (§3); more massive clusters attract each other more strongly than less massive clusters on small scales ($r \leq 10h^{-1}\text{Mpc}$).

### 4.2 RMS Pairwise Velocity Dispersion of Clusters

The 1D rms pairwise peculiar velocity dispersion of clusters, $\sigma_{12}^{1D} \equiv \langle(\vec{v_1} - \vec{v_2})^2\rangle^{1/2}/\sqrt{3}$, is presented as a function of pair separation, $r$, for all models, in Figure 4. Unlike $v_{12}(r)$ discussed in §4.1, $\sigma_{12}^{1D}$ does not determine the mean relative direction of the pairs motion; rather, it provides the average magnitude of the rms velocity dispersion of cluster pairs.

The cluster velocity dispersion exhibits large pairwise velocities on small scales ($\sim 400$ to $1300\text{km s}^{-1}$ for $r \leq 10h^{-1}\text{Mpc}$, depending on the model); the velocity dispersion decreases nearly monotonically with scale, similar to the $v_{12}(r)$ function, and reaches a large-scale velocity dispersion of $\sim 300\text{km s}^{-1}$ (LCDM) and $\sim 600\text{km s}^{-1}$ (SCDM and PBI) on $\sim 100h^{-1}\text{Mpc}$ scale. All models exhibit a similar behavior, but have different amplitudes (Figure 3 and Table 1). The broad minimum near $30h^{-1}$Mpc reflects the transition beyond which the bulk motions become small relative to the overall velocities.

Figure 3. Cluster pairwise velocity difference as a function of pair separation for the three models (§4.1). ($R \geq 1$ clusters).

The large velocity dispersion on small scales reflects the strong mean attraction of clusters on these scales, as implied from $v_{12}$ (§4.1). The ratio $v_{12}/\sigma_{12}^{1D}$

Figure 4. Cluster 1D pairwise rms velocity dispersion as a function of pair separation for the three models (§4.2). ($R \geq 1$ clusters).

is nearly $\sqrt{3}$ at $r \sim 5h^{-1}$Mpc, as expected if the cluster pairwise motions on these scales are mostly one-dimensional attractive motions along the line joining the cluster pair; in that case $v_{12} \sim \sigma_{12}^{3D} \sim \sqrt{3}\sigma_{12}^{1D}$, as observed. This attractive motion is strongest for SCDM, with $\sigma_{12}^{1D}(5h^{-1}Mpc) \sim 1000$km s$^{-1}$, and is lower for PBI ($\sim 650$km s$^{-1}$) and LCDM ($\sim 500$km s$^{-1}$). The results are summarized

– 9 –

in Table 1. The large difference in the cluster pairwise velocity dispersion on these scales makes this parameter an important tool in constraining cosmological models when compared with observations.

The decrease of the cluster velocity dispersion with scale to $r \sim 10h^{-1}$Mpc, and its flattening at a constant positive dispersion on large scales, reflects the transition from the gravitational attraction regime of close cluster pairs, to the overall rms pairwise velocities of clusters on large scales (Fig. 4, Table 1). Cluster pair velocities are mostly uncorrelated on large scales. In this case $\sigma_{12}^{1D} \sim \sqrt{2} v_{rms}^{1D}$, where $v_{rms}^{1D}$ is the 1D rms velocity of clusters. This relation is indeed satisfied: $v_{rms}^{1D}$ of rich clusters is 446, 254, and 411km s$^{-1}$ for SCDM, LCDM, and PBI models, respectively (Gramann et al. 1994); these values are $\sim \sqrt{2}$ lower than the large-scale $\sigma_{12}^{1D}$ ($100h^{-1}$Mpc) dispersions listed in Table 1.

The ratio $v_{12}/\sigma_{12}^{1D}$ decreases monotonically with scale from $\sqrt{3}$ on small scales to zero at $r \sim 100h^{-1}$Mpc. The trend is similar in all models. This behavior is expected if the cluster pairwise motions change from one-dimensional attraction on small scales to bulk and finally random motions on large scales [where $v_{12}$ decreases to zero, and $\sigma_{12}^{1D}$ to a constant, as observed].

The amplitude of the cluster pairwise velocity dispersion on small scales ($r \leq 10h^{-1}$Mpc) depends on cluster mass; more massive clusters yield larger pairwise velocity dispersions, especially for SCDM. The richness (mass) dependence is consistent with that observed for the velocity correlation function (§3) and $v_{12}(r)$ (§4.1); it suggests stronger attractive motions for more massive clusters on small scales. The large scale motions do not depend on cluster mass.

## 5. CONCLUSIONS

We determine the velocity correlation function of rich clusters of galaxies, and the pairwise peculiar velocity difference and rms pairwise peculiar velocity dispersion of clusters for three cosmological models: $\Omega = 1$ and $\Omega = 0.3$ CDM, and $\Omega = 0.3$ PBI. Our principal conclusions are listed below.

1. All models show that close cluster pairs ($r \leq 10h^{-1}$Mpc) exhibit strong attractive velocities. The mean pairwise attractive velocity on $5h^{-1}$Mpc scale ranges from $v_{12}(5h^{-1}\text{Mpc}) \sim 1700$km s$^{-1}$ for $\Omega = 1$ CDM to $\sim 700$km s$^{-1}$ for $\Omega = 0.3$ CDM. The pairwise cluster velocity function decreases monotonically with scale, reflecting the relative decrease of the pairwise cluster gravitational attraction and the relative increase of random and bulk motions on large scales. The random and bulk motions finally dominate on large scales, $r \sim 50 - 100h^{-1}$Mpc, where $v_{12}(r) \sim 0$. Massive clusters exhibit somewhat stronger attractive velocities than less massive clusters on small scales; on large scales, where the cluster velocities are dominated by bulk and random motions, the velocities are independent of cluster richness.

2. The rich cluster velocity correlation function reveals positive correlations to very large scales, from $r \sim 10h^{-1}$Mpc to $200h^{-1}$Mpc, for all models (Fig 1). The positive correlations decrease monotonically with scale for $r \geq 20h^{-1}$Mpc, approaching a nearly random distribution of cluster velocities at $r > 200h^{-1}$Mpc.

The positive correlations on $\sim 20 - 200h^{-1}$Mpc scale reflect bulk motions of clusters, produced by large-scale gravitational perturbations. On small scales, $r \leq 10h^{-1}$Mpc, the velocity correlations are negative for $\Omega = 1$ and $\Omega = 0.3$ CDM, reflecting strong mean attractive motions for close cluster pairs, especially for $\Omega = 1$. On scales up to $10^2 h^{-1}$Mpc, PBI exhibits the largest positive velocity correlations of clusters, reflecting strong bulk motions.

3. The ratio $v_{12}/\sigma_{12}^{1D}$ decreases with scale from $\sim \sqrt{3}$ on small scales ($\sim 5h^{-1}$Mpc) to zero on large scales. This demonstrates that clusters in close pairs move primarily along the line joining the pair while pairs with large separations move mostly randomly.

4. The rich cluster pairwise velocity functions and velocity correlation function discussed above depend strongly on the cosmological model (shape and amplitude of the power spectrum and $\Omega$). These functions (or some variants of them) can therefore serve as a potentially powerful tool in constraining cosmological models when compared with observations.

It is a pleasure to acknowledge NCSA for use Convex-3880 supercomputer. Support by NASA grant NAGW-2448, NSF grants AST90-20506, AST91-08103, AST93-15368 and HPCC ASC-9318185 are gratefully acknowledged.

TABLE 1

VELOCITY CORRELATIONS AND PAIRWISE VELOCITIES

OF RICH (R $\geq$ 1) CLUSTERS*

| Model | $\psi_v(5)$ | $\psi_v(20)$ | $\psi_v(100)$ | $v_{12}(5)$ | $\sigma_{12}^{1D}(5)$ | $\sigma_{12}^{1D}(100)$ |
|---|---|---|---|---|---|---|
| CDM $\Omega = 1.0$ | -692 | 426 | 235 | 1698 | 1070 | 595 |
| CDM $\Omega = 0.3$ | -87 | 275 | 162 | 714 | 487 | 327 |
| PBI $\Omega = 0.3$ | 451 | 554 | 225 | 962 | 646 | 555 |

* The velocities are in km s$^{-1}$. The scale at which each parameter is listed, indicated inside the parentheses, is in units of $h^{-1}$Mpc. The statistical pair uncertainties are approximately 15% at $5h^{-1}$Mpc, 5% at $20h^{-1}$Mpc, and lower at large scales.